\documentclass[preprint,12pt, a4paper]{elsarticle}

\usepackage{amssymb}
\usepackage{amsmath}
\usepackage{tensor}
\usepackage{hyperref}
\usepackage{url}
\usepackage{xcolor}
\usepackage{romannum}
\usepackage{xspace}
\usepackage[ruled,vlined]{algorithm2e}
\usepackage{array}
\usepackage{soul}
\usepackage[normalem]{ulem}

\newcommand{\hard}{\textsc{HARD}\xspace}
\newcommand{\flecsi}{\textsc{FleCSI}\xspace}
\newcommand{\flecsolve}{\texttt{FleCSolve}\xspace}

\journal{SoftwareX}

\begin{document}
\renewcommand{\labelenumii}{\arabic{enumi}.\arabic{enumii}}

\begin{frontmatter}

\title{HARD: A Performance Portable Radiation Hydrodynamics Code based on \flecsi Framework}


\author[label1]{Julien Loiseau}
\author[label1]{Hyun Lim}
\author[label1]{Andrés Yagüe López}

\author[label2]{Mammadbaghir Baghirzade}
\author[label3]{Shihab Shahriar Khan}
\author[label4]{Yoonsoo Kim}
\author[label5]{Sudarshan Neopane}
\author[label6]{Alexander Strack}
\author[label7]{Farhana Taiyebah}
\author[label1]{Ben Bergen}

\address[label1]{Los Alamos National Laboratory}
\address[label2]{The University of Texas at Austin}
\address[label3]{Michigan State University}
\address[label4]{Princeton Center for Theoretical Science}
\address[label5]{University of Tennessee, Knoxville}
\address[label6]{Universität Stuttgart}
\address[label7]{Florida State University}

\begin{abstract}

\textbf{Hydrodynamics And Radiation Diffusion} (\hard) is an open-source
application for high-performance simulations of compressible hydrodynamics with
radiation-diffusion coupling. Built on the \flecsi~\cite{bergen2021flecsi}
(Flexible Computational Science Infrastructure) framework, \hard expresses its
computational units as tasks whose execution can be orchestrated by multiple
back-end runtimes, including Legion~\cite{bauer2012legion},
MPI~\cite{mpi1994standard}, and HPX~\cite{Kaiser2020hpx}. 
Node-level parallelism is delegated to Kokkos~\cite{edwards2014kokkos}, providing a single, portable code base that runs efficiently on laptops, small homogeneous clusters, and the largest heterogeneous supercomputers currently available.

To ensure scientific reliability, \hard includes a regression-test suite that
automatically reproduces canonical verification problems such as the Sod and
LeBlanc shock tubes, and the Sedov blast wave, comparing numerical solutions
against known analytical results. The project is distributed under an
OSI-approved license, hosted on GitHub, and accompanied by reproducible build
scripts and continuous integration workflows. This combination of performance
portability, verification infrastructure, and community-focused development
makes \hard a sustainable platform for advancing radiation hydrodynamics
research across multiple domains. \end{abstract}


\end{frontmatter}


\section*{Metadata}
\label{}

\begin{table}[!ht]
\begin{tabular}{|l|p{6.5cm}|p{6.5cm}|}
\hline
\textbf{Nr.} & \textbf{Code metadata description} & \textbf{Metadata} \\
\hline
C1 & Current code version & V1.0 \\
\hline
C2 & Permanent link to code/repository used for this code version & \url{https://github.com/lanl/hard} \\
\hline
C3  & Permanent link to Reproducible Capsule & \\
\hline
C4 & Legal Code License   & BSD 3-Clause License \\
\hline
C5 & Code versioning system used & git \\
\hline
C6 & Software code languages, tools, and services used & C++, MPI \\
\hline
C7 & Compilation requirements, operating environments \& dependencies & Kokkos, Singularity-EOS \\
\hline
C8 & If available Link to developer documentation/manual & \url{https://lanl.github.io/HARD/} \\
\hline
C9 & Support email for questions & hard-help@lanl.gov\\
\hline
\end{tabular}
\caption{Code metadata (mandatory)}
\label{codeMetadata} 
\end{table}

\section{Motivation and significance}

Radiation hydrodynamics (RHD) governs systems where radiation–matter coupling
shapes the global dynamics, including high-energy-density physics (HEDP) problems such as inertial confinement fusion (ICF), stellar evolution and explosions, accretion flows, and radiative shocks. Accurately capturing these phenomena requires
solving tightly coupled, nonlinear partial differential equations for hydrodynamics and radiation
diffusion across disparate spatial and temporal scales, often in the presence of
stiff source terms and strong gradients.

While several mature RHD and radiation–Magnetohydrodynamics (MHD) codes
exist~\cite{Turner2001,vanderHolst:2011md,Zhang:2011ev,Jiang:2022ccq,Moens2022,Wibking2022},
many legacy implementations face practical challenges on modern heterogeneous
platforms such as limited performance portability, rigid data models that impede
algorithmic experimentation, and difficulty integrating with contemporary
tooling for CI, packaging, and reproducibility. As experimental capabilities and
computing architectures evolve, a software platform that cleanly separates
physics from execution back ends, and that scales from laptops to
leadership-class systems, becomes essential.

\hard is designed to meet this need. Built on the task-based \flecsi runtime
layer, it decouples application physics from parallel orchestration. This design
enables the same source to target multiple distributed-memory runtimes (Legion,
MPI, and HPX) while delegating node-level parallelism to Kokkos through an abstraction layer, yielding a single, portable implementation for CPUs and GPUs. Beyond portability, \hard
aims to serve as a \emph{testbed} for new methods in RHD by providing clear
interfaces for numerics (reconstruction, Riemann solvers, time integrators) and
physics (equations of state, radiation closures).

To promote scientific reliability and re-use, \hard includes:
\begin{itemize}
  \item a regression test suite that automatically reproduces canonical verification problems (e.g., Sod and LeBlanc shock tubes, Sedov blast) and compares against analytic or reference solutions;
  \item containerized and scriptable builds to ensure bitwise-reproducible environments across developer machines and CI;
  \item open-source governance and contribution workflows to lower the barrier for community extensions (e.g., MHD, multigroup radiation, advanced microphysics).
\end{itemize}

In summary, \hard addresses two complementary gaps: (i) a performance-portable
RHD code that maps cleanly to heterogeneous architectures, and (ii) a modular,
reproducible platform that accelerates method development and cross-code
comparison for the RHD community.

\section{Software description}
\label{sec:sw_description}

\hard is implemented in modern C++ for UNIX and high-performance computing (HPC)
platforms, taking advantage of the Standard Template Library (STL). Spack~\cite{Gamblin2015} is
used to configure both build- and run-time environments. Simulation data is
written in CSV and Catalyst formats, enabling post-processing with visualization
tools such as ParaView~\cite{Ayachit2015} and VisIt~\cite{Childs2012}.  

\subsection{Software architecture}
\label{sec:sub:sw_arch}

\begin{figure}[h!]
\centering
\includegraphics[width=0.7\textwidth]{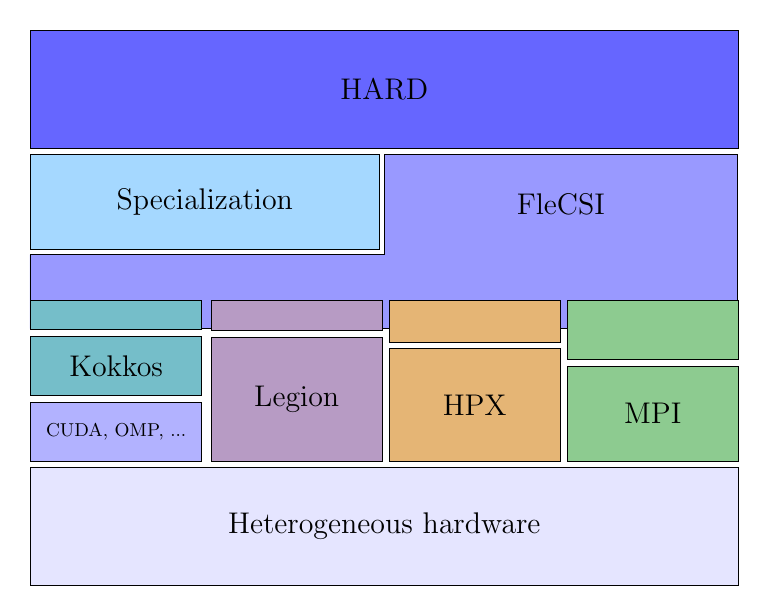}
\caption{\hard within the \flecsi ecosystem.}
\label{hard_diagram}
\end{figure}

The software stack is shown in Figure~\ref{hard_diagram}. \hard uses a \flecsi
specialization layer that employs the N-Array topology (a multidimensional array
with exclusive, shared, and ghost entities).

The \hard base code defines its specialization in \texttt{spec/}, and the
application that uses the specialization in \texttt{app/}. The current control
model consists of initialization, time-stepping (advection, diffusion, and
source updates), and output. 

\subsection{Software functionalities}
\label{sec:sub:sw_func}

\hard provides a set of core capabilities for solving RHD
problems on modern HPC systems. At a high level, these include:
\begin{itemize}
    \item implementation of the coupled hydrodynamics and radiation–diffusion equations,  
    \item modular numerical algorithms (finite-volume solvers, reconstruction methods, Riemann solvers, and implicit diffusion solvers),  
    \item verification problems and regression tests for scientific reliability, and  
    \item scalable distributed parallel execution across CPUs and GPUs via \flecsi back ends.  
\end{itemize}

The following subsections describe the mathematical models, numerical
algorithms, and example problems that illustrate these capabilities.

\subsubsection{Basic Equations}
\label{sec:sub:sub:eqns}

We adopt the co-moving (Eulerian) frame formulation of the RHD equations in the
diffusion limit~\cite{Mihalas1984}. This formulation is particularly well-suited
for regimes where the mean free path of photons is small compared to the
characteristic hydrodynamic length scales, allowing the radiative flux to be
described by a diffusion approximation. In this framework, the governing
equations can be compactly expressed in Einstein notation as
\begin{equation} \label{eq:conservative form}
   \partial_t \mathbf{U} + \partial_j \mathbf{F}^j = \mathbf{S}
\end{equation}
with the evolved variables
\begin{equation}
   \mathbf{U} = \left[ \rho, \rho v^i, e + \rho v^2 / 2, E \right]^T,
\end{equation}
fluxes
\begin{equation} \label{eq:fluxes}
    \mathbf{F}^j = \left[ \begin{matrix}
        \rho v^j \\
        \rho v^i v^j + p \delta_{ij} \\
        \left(e + \rho v^2/2 + p\right) v^j \\
        E v^j \\
        \end{matrix} \right],
\end{equation}
and source terms
\begin{equation}
   \mathbf{S} = \left[ \begin{matrix}
   0 \\
   f^i \\
   f^i v_i + \dot{q} \\
   - \partial_i F_r^i - P_r^{ij}\nabla_i v_j - \dot{q} \\
   \end{matrix} \right],
\end{equation}
where $\rho$ is fluid mass density, $v^i$ is the component $i$ of the fluid
velocity vector, $e$ is the fluid internal energy density, $p$ is the fluid
pressure, $E$ is radiation energy density, $F_r^i$ is the component $i$ of the
radiation energy flux vector, $P_r^{ij}$ is the $i$, $j$ element of the
radiation pressure tensor, $f^i$ is the component $i$ of the radiation force
density vector
\begin{equation}
   f^i = \frac{\kappa \rho}{c} F_r^i ,
\end{equation}
and $\dot{q}$ is the radiative heating rate
\begin{equation}
\label{eqn:heating-cooling}
   \dot{q} = \frac{dq}{dt} = c \kappa \rho (E - a T^4) ,
\end{equation}
where $\kappa$ is the mean opacity, $c$ is the speed of light, $T$ is the fluid
temperature, and $a$ is the radiation defined as $a\equiv4\sigma/c$, where $\sigma$
is the Stefan-Boltzmann constant.

To close the evolution system for the fluid, it is necessary to specify the
functional form of the fluid pressure $p$, which can be obtained from an
equation of state (EoS). In this work, we adopt
\texttt{Singularity-EOS}~\cite{singularity-eos}, which provides a flexible EoS
framework that enables the modeling of different materials. The radiative
transfer equation—the final component of the evolution system given in
Equation~\ref{eq:conservative form}—further requires closure relations for both
the radiation energy flux density $F^i$ and the radiation pressure tensor
$P_r^{ij}$.

In the optically-thick diffusion limit, they are given as
\begin{equation}
    F^i = - \frac{c}{3\kappa\rho} \nabla E,
\end{equation}
\begin{equation}
    P_r^{ij} = \frac{1}{3} E \delta_{ij},
\end{equation}
provided the mean free path of photons are much shorter than the length scale of interest.

The evolution equations presented above are valid in the diffusion (optically
thick) limit. However, it is possible to extend their applicability to optically
thin regimes by introducing a bridge law~\cite{Turner2001}, which ensures that
the radiation energy flux density $F_r^i$ attains the correct magnitude in the
free-streaming limit.
\begin{equation}
   F_r^i = - \lambda \frac{c}{\kappa \rho} \nabla E
\end{equation}
Here, $\lambda$ is a radiation energy flux limiter function~\cite{Levermore1981}
\begin{equation}
   \lambda = \frac{2 + R}{6 + 3R + R^2}
\end{equation}

\begin{equation}
   R \equiv \frac{|\nabla E|}{\rho \kappa E}
\end{equation}

\begin{equation}
   \frac{P^{ij}}{E} = \left(\frac{1-f}{2}\right) \delta_{ij}
      + \left(\frac{3f-1}{2}\right) n^i n^j
\end{equation}
where $n^i \equiv \nabla E/|\nabla E|$ and the quantity $f$ is defined as
\begin{equation}
   f = \lambda + \lambda^2 R^2 .
\end{equation}

In the optically thick limit, where $R \to 0$, one finds $\lambda \to 1/3$ and
$f \to 1/3$, such that the Eddington approximation, $P^{ij} = (E/3)\delta_{ij}$,
is recovered. In contrast, in the optically thin limit, the photon field streams
along the gradient of the local radiation energy density at the speed of light.
Although these prescriptions do not constitute a fully self-consistent treatment
of light–matter interactions, they provide a computationally efficient closure
that smoothly bridges the optically thick and thin regimes. As a result, the
radiative transfer equation becomes

\begin{equation}\label{eq:radiation}
    \partial_t E + \partial_j (Ev^j) =
        \partial_i(D\, \partial_i E) - P_r^{ij}\nabla_i v_j - \dot{q} ,
\end{equation}
with $D = c\lambda / \kappa\rho$.

\subsubsection{Numerical algorithms}
\label{sec:sub:sub:num_algo}

The computational domain is represented by a uniform Cartesian mesh, with
evolved variables located at the center of each cell. In this section, we
abandon Einstein notation and limit our notation to a single dimension for
clarity.

We use an operator split approach to handle the conservative advection equations
(\ref{eq:fluxes}) and the non-advective part of the radiation transfer equations
(\ref{eq:radiation}).

The advection parts are solved using a finite volume method, with the
discretization

\begin{equation}
    \frac{\partial F_x}{\partial x} \to \frac{F^*_{x_{i+1/2}} - F^*_{x_{i-1/2}}}{\Delta x_i}
\end{equation}

where $F^*_{x_{i\pm 1/2}}$ are the numerical fluxes computed at each of the cell
faces $x_{i\pm 1/2}$ around cell coordinate $x_i$ in meshpoint $i$, and are
obtained with an approximate Riemann solver along with an appropriate face
reconstruction heuristic from cell-centered values. We employ the WENO5-Z
reconstruction~\citep{Borges2008} for all variables, where we adopt the HLL
fluxes~\citep{Harten1983} for hydrodynamics variables and the local
Lax-Friedrichs (Rusanov) fluxes~\citep{Rusanov1962} for the radiation energy
density. The time advance is done using Heun's method for a second-order
approximation in time. A Butcher tableau is given in
Table~\ref{tab:heuns_butcher}.

\begin{table}[h]
    \centering
    \begin{tabular}{c | c c}
         0 &  & \\
         1 & 1 & \\
         \hline  
         & 1/2 & 1/2
    \end{tabular}
    \caption{Butcher tableau for Heun’s method (second-order RK).}
    \label{tab:heuns_butcher}
\end{table}

The diffusive part of the radiative transfer equations is solved with a
geometric multigrid solver (GMG) with regular coarsening. GMG solves the
elliptic equation

\begin{equation}\label{eq:linear_equation}
    Au = f\,,
\end{equation}

where $A$ is a linear operator. Multigrid solvers use successively coarser grid
representations of the simulation domain to accelerate convergence by a
recursion that combines the ability of elementary iterative solvers to damp
high-frequency errors with coarse-grid error corrections. On each coarse grid
level, elementary iterative solvers are again effective at damping
high-frequency error until some reduced set of unknowns can be solved directly.
The resulting error corrections are prolonged to the next finer grid with some
small number of smoothing cycles applied to eliminate noise. Using this
technique appropriately results in asymptotically optimal convergence, i.e.,
O(n) for n unknowns~\cite{Briggs2000, Trottenberg2000}.

For every grid except for the so-called ``solution grid'' (i.e., the finest
grid), the equation being solved is the residual of Equation
(\ref{eq:linear_equation}),

\begin{equation}\label{eq:residual_equation}
    r = f - Au' \implies Ae = r\,,
\end{equation}

with $e = u_c - u'$ where $u_c$ is a better approximation to the solution $u$
given by the previous approximation $u'$. The advantage of the residual is that
a good initial guess is well known as $r_0 = 0$. The error field $e$ is then
transported to the finer grid to correct the previous approximation to the
solution $u'$ in that level. Equation (\ref{eq:residual_equation}) is only
correct if

\begin{equation}
    Ae = r \implies A(u_c - u') = f - Au' \implies Au_c = f\,,
\end{equation}

which necessitates that $A$ is a linear operator.

The coarsest level in this progression ideally has many fewer nodes than the
solution grid allowing for computationally expensive but more accurate methods,
such as the Conjugate Gradient method. In our current implementation of the code
a simple relaxation method is used instead. In the future, we will use linear
solvers from the parallel solver library \flecsolve~\cite{flecsolve} for this
step.

\begin{figure}[h!]
\centering
\includegraphics[width=0.7\textwidth]{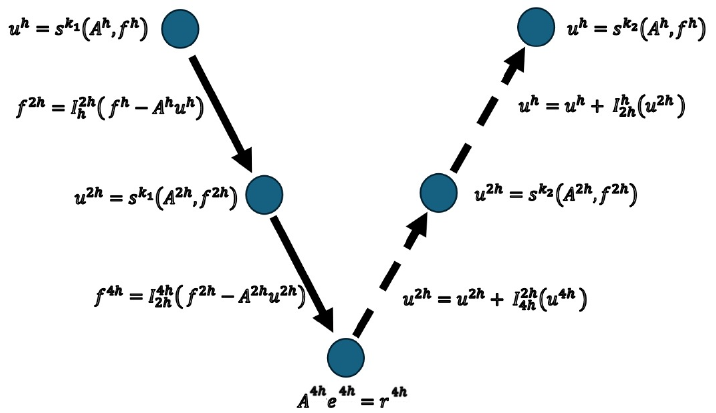}
\caption{A v-cycle, the most basic structure of a GMG solver. Every circle represents a different grid level, with circles higher up representing finer grids where $h$ is the distance between two consecutive nodes in the grid. The restriction and prolongation operations are represented with the $I$ operator, and the relaxation solver applied $k$ times is represented with the $s^k$ operator. The specific notation used in this figure can be found in any of the references provided for GMG.}
\label{fig:v-cycle}
\end{figure}

The most basic structure of a GMG solver is known as a v-cycle, illustrated in figure~\ref{fig:v-cycle}.

\section{Illustrative examples}
\label{sec:examples}

In this section, we provide several cases to demonstrate validation and
functionalities of the code. The problems used to test the code fall into three
categories. First, we demonstrate with hydrodynamics problem, then we isolate
the heating and cooling term, and finally show the full RHD test problems. 

For the following tests, we will quantify the error between the numerical and
analytical solutions with the $L_1$ metric defined as

\begin{equation}
  L_1 = \Delta V\sum\left|N_i - \langle A(x_i)\rangle \right|,
\end{equation}\label{eq:l1_error}

where $N_i$ is the numerical solution at grid coordinate $x_i$, $\langle
A(x_i)\rangle$ is the volume-averaged analytical solution over the finite volume
containing element $i$, and $\Delta V$ is the volume element of the regular
grid. $\langle A(x_i)\rangle$ is calculated using a Gauss-Legendre quadrature
with 10 nodes.

\subsection{Basic hydrodynamics problem: Sod shock tube test}
\label{sec:sub:sod}

The Sod shock tube is a standard Riemann problem test with the following initial
parameters:
\begin{equation}
(\rho, v, p)_{t=0} = 
\begin{cases}
(1.0,0.0,1.0) & \text{if} \,\, 0.0 < x \leq 0.5, \\
(0.125,0.0,0.1) & \text{if} \,\, 0.5 < x < 1.0.
\end{cases}
\end{equation}
This leads to the development of a shock front, which propagates from
high-density into low-density regions, followed by a contact discontinuity,
while a rarefaction wave propagates.

\begin{figure}
    \centering
    \includegraphics[width=0.7\linewidth]{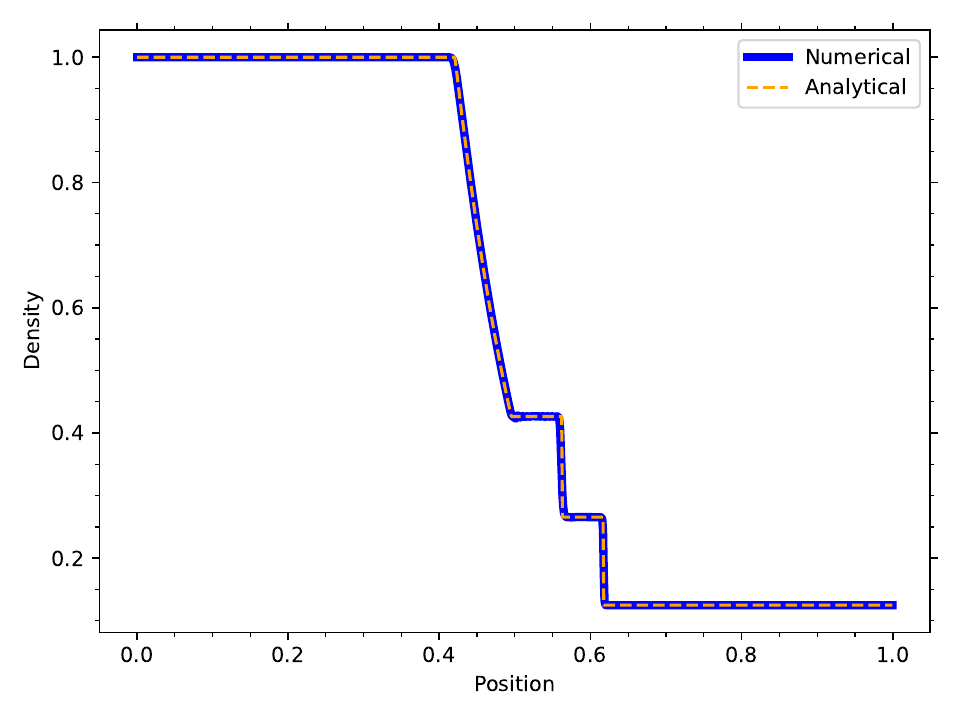}
    \caption{Numerical and analytical solution to the Sod problem at simulation time $t = 0.0669$ with $dx = 2^{-10}$ and $dt = 0.3$ CFL. The $L_1$ error for this figure is below $5\cdot10^{-4}$.}
    \label{fig:sod_density}
\end{figure}

The numerical and analytical solution comparison for density in the Sod tube in
HARD is shown in Figure~\ref{fig:sod_density}.

\subsection{Radiation test problem: heating and cooling}
\label{sec:sub:hc}

We perform a standard heating and cooling of the fluid by
radiation~\cite{Roth2015}. The mass density of the fluid is set to $\rho =
10^{-7}\,\mathrm{g\,cm^{-3}}$, with a mean molecular weight of $\mu = 0.6$ and an
opacity of $\kappa = 0.4\,\mathrm{cm^2\,g^{-1}}$. A uniform radiation field is
initialized with a temperature of $T_{0,\mathrm{r}} = 3.4 \times
10^6\,\mathrm{K}$. In this test, we consider two scenarios: the fluid is heated
by radiation; and the fluid cools. For the heating case, the fluid is
initialized at $T_{0,f,\mathrm{heating}} = 11\,\mathrm{K}$, while for the cooling
case the initial temperature is $T_{0,f,\mathrm{cooling}} = 1.1 \times
10^9\,\mathrm{K}$. In both scenarios, the radiation energy density exceeds the
internal energy of the fluid and, thus, remains nearly constant during the
energy exchange. Consequently, the fluid eventually equilibrates to the
radiation temperature.


An analytic benchmark solution can be obtained by solving the ODE in Equation
(\ref{eqn:heating-cooling}).

\begin{figure}
    \centering
    \includegraphics[width=0.7\linewidth]{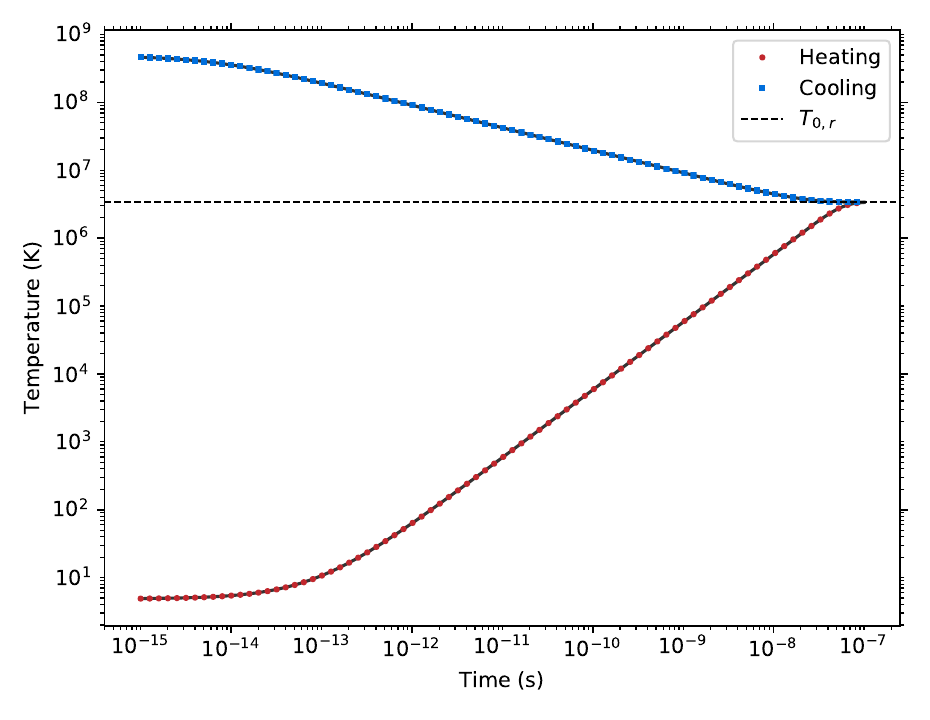}
    \caption{Tests of the approach to radiative equilibrium in a radiation dominated gas. Both cases approach the equilibrium value and matches with the analytical solution (solid lines).}
    \label{fig:heating_cooling}
\end{figure}

Figure~\ref{fig:heating_cooling} presents the numerical results of the
heating–cooling test alongside a reference solution (solid lines) obtained by
integrating Equation (\ref{eqn:heating-cooling}) with an ODE solver. The numerical
solutions show agreement with the reference results and asymptotically converge
to the radiation temperature, as expected.

\subsection{Radiation hydrodynamics problem: temperature induced shock}
\label{sec:sub:tshock}

In scenarios such as ICF, where energy from laser or X-ray drivers rapidly heats
a thin outer layer of a fusion capsule, thermal energy deposition creates steep
pressure gradients that launch shock waves. In these systems, radiation
diffusion plays a critical role by preheating material ahead of the shock front,
altering the shock structure and modifying compression and temperature profiles.
To test this phenomenon, we perform temperature-induced shock problems in
1D~\cite{Rage_man}. 
\begin{figure}
    \centering
    \includegraphics[width=0.7\linewidth]{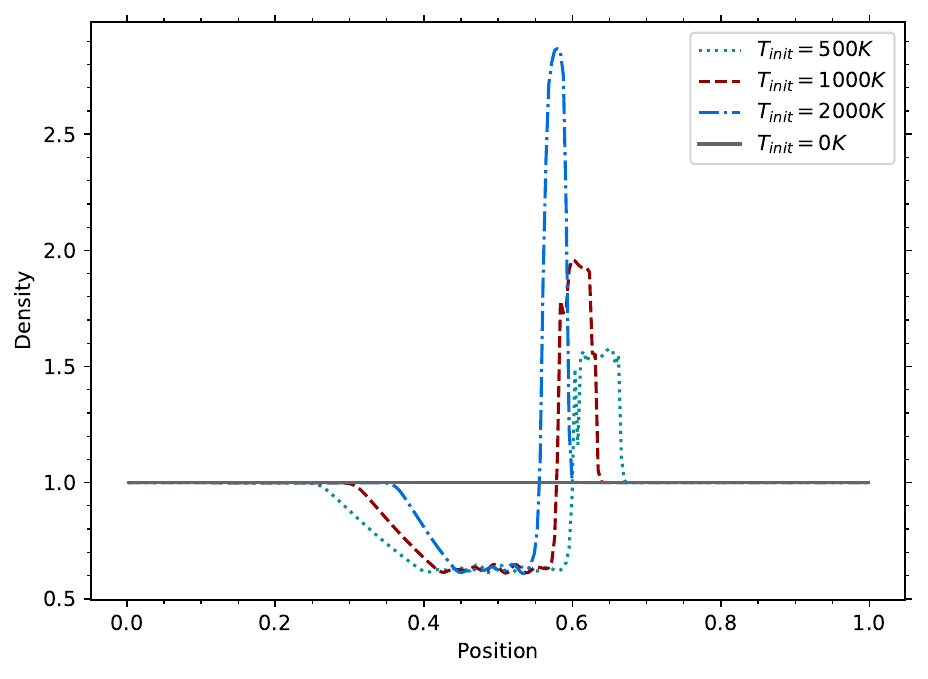}
    \caption{The Temperature induced shock test problem with different initial temperatures at $t=0.5$.
    The radiation temperature is the only driving source, so $T=0$ K shows no shock formation. In contrast, nonzero-$T$ cases exhibit shock formation.}
    \label{fig:temp_shock}
\end{figure}
Figure~\ref{fig:temp_shock} shows a temperature-induced shock test problem that is
examined at time $t=0.5$ for different initial temperatures. We apply an ideal EoS
for this test problem. Since radiation temperature is the sole driving
mechanism, the case with $T=0K$ does not exhibit shock formation. In contrast,
cases with nonzero initial temperatures demonstrate the development of shock
structures. This test provides a simplified yet powerful framework to benchmark
simulation codes, explore radiation–matter coupling, and study shock timing and
strength, all essential for optimizing performance in ICF experiments.

\subsection{Radiation hydrodynamics problem: radiative Kelvin-Helmholtz instability}
\label{sec:sub:kh}

The Kelvin-Helmholtz instability (KHI) is a fundamental shear-flow instability that
arises when there is velocity shear within a continuous fluid or across the
interface between two fluids moving at different
speeds~\cite{Chandrasekhar1961}. KHI is important to understand various
phenomena, from shear layers in accretion flows and at stellar surfaces to
laboratory experiments relevant to inertial confinement fusion.

\begin{figure}
    \centering
    \includegraphics[width=0.49\linewidth]{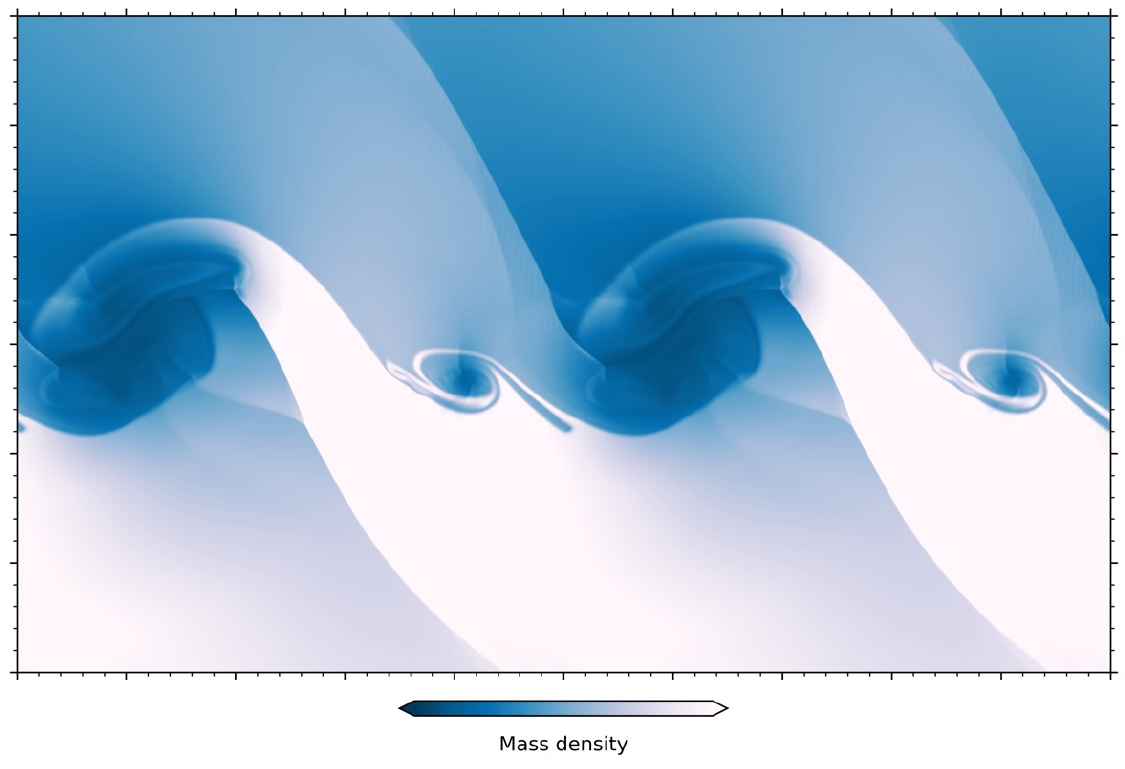}
    \includegraphics[width=0.49\linewidth]{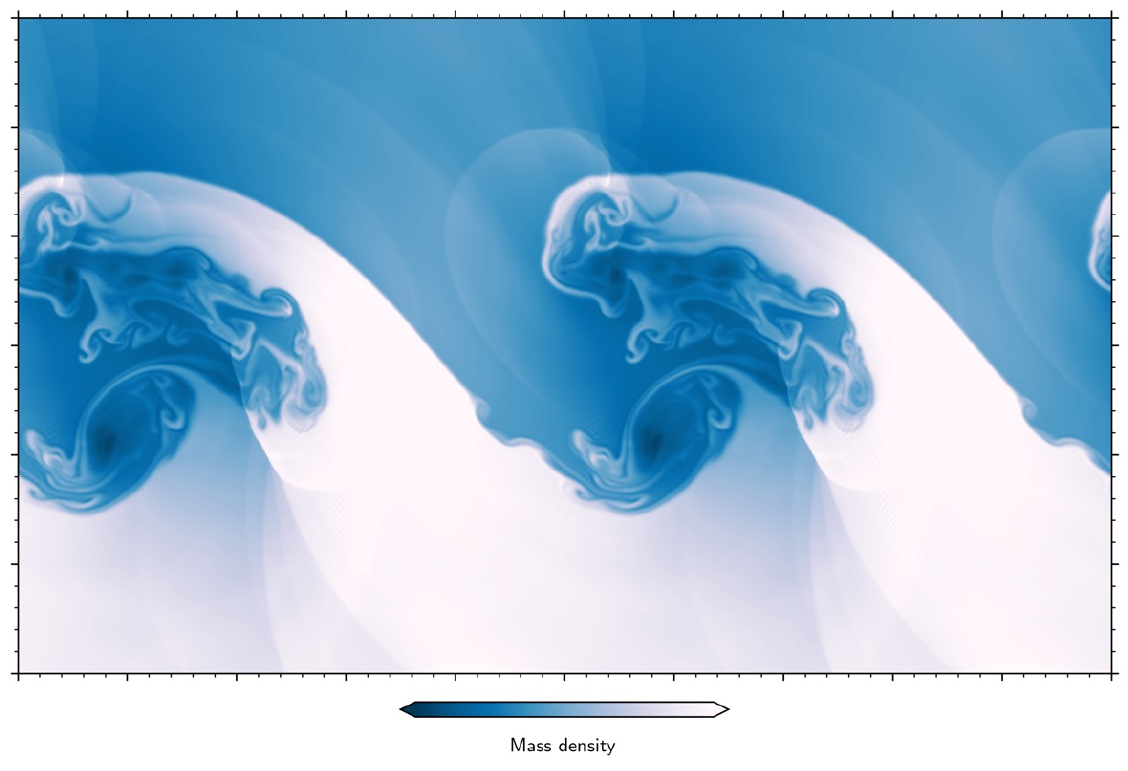}
    \caption{Density plot of Kelvin-Helmholtz instability with (right) and without (left) radiation at $t=16s$. The radiation case shows the instability is destabilizing with the presence of radiation field. }
    \label{fig:rad_KH}
\end{figure}
Linear analyses~\cite{Shadmehri2012,Peng2024} show that the presence of a
radiation field increases the growth rates of unstable modes, thereby shortening
the characteristic instability timescale. Motivated by this result, we perform
two-dimensional simulations of KHI with and without radiation. The computational
domain is the square $\{(x,y)\,|\, -0.5 \le x \le 0.5,\; -0.5 \le y \le 0.5\}$.
We initialize two layers of different density, $\rho_{\mathrm{low}}=1.0$ and
$\rho_{\mathrm{high}}=2.5$, and prescribe a piecewise-constant streamwise
velocity: $v_x=-0.5$ for $|y|\le 0.25$ and $v_x=0.5$ elsewhere. In the radiative
case, a spatially uniform radiation temperature
$T_{\mathrm{rad}}=1000\,\mathrm{K}$ is imposed throughout the domain.

Figure~\ref{fig:rad_KH} shows the density field at $t=16\,\mathrm{s}$ for the
radiative (right) and non-radiative (left) runs. Consistent with linear theory,
radiation accelerates the development of KHI, leading to faster growth of
instabilities. A systematic exploration of parameter dependence and radiative
effects will be reported in future work.

\subsection{Numerical Convergence}
\label{sec:sub:conv}

\begin{figure}[ht!]
    \centering
    \includegraphics[width=0.7\linewidth]{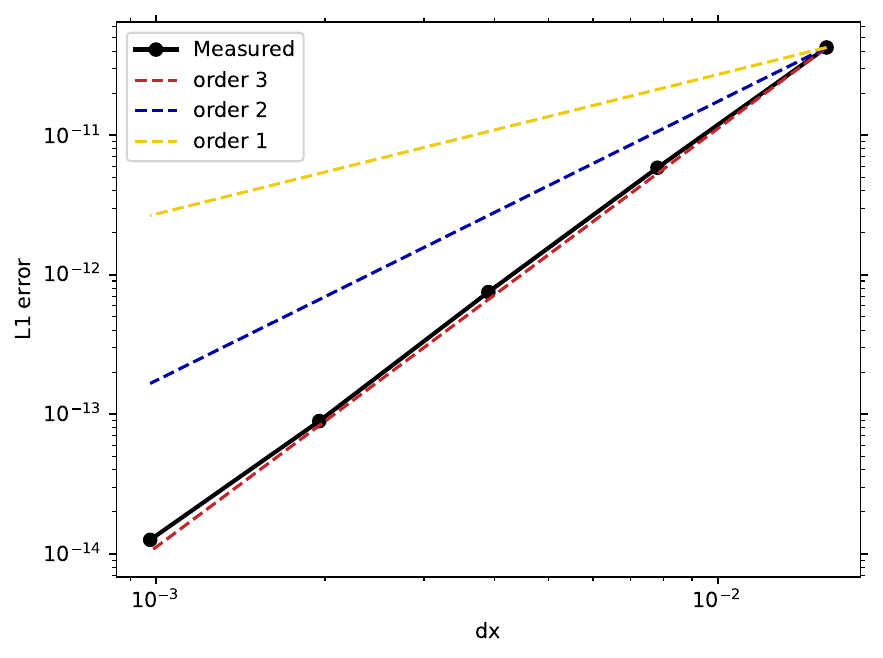}
    \caption{Convergence order for an acoustic wave problem with weno5z as limiter}
    \label{fig:convergence}
\end{figure}

To study the convergence of the code, see Figure~\ref{fig:convergence}, where we run
an acoustic wave problem with a WENO5-Z limiter~\cite{Turner2001}. We choose the
same final time for subsequent, smaller grids, with a timestep given by the
Courant–Friedrichs–Lewy (CFL) condition.

The acoustic wave problem can be understood by linearizing
Equation (\ref{eq:conservative form}), defining each primitive field as a
homogeneous equilibrium value and a \textit{small} perturbation. Assuming the
perturbation shape can be decomposed in Fourier modes, it is enough to apply the
linearized equations to one such Fourier mode. From this we obtain the following
system of algebraic equations:
\begin{equation}
    \begin{matrix}
        \omega \frac{\rho_A}{\rho_0} = \mathbf{k}\cdot\mathbf{u_A} \\
        \omega \mathbf{u_A} = \gamma\frac{P_0}{\rho_0}\frac{\rho_A}{\rho_0}\mathbf{k}, \\
        \end{matrix}
\end{equation}\label{eq:acoustic_algebraic}
which relate the frequency of the mode $\omega$, its wave number $\mathbf{k}$,
the equilibrium values for pressure $P_0$ and density $\rho_0$, and the
perturbation amplitudes for density $\rho_A$ and velocity $\mathbf{u_A}$. Here
the acoustic waves are assumed to be an adiabatic process in an ideal gas, so
the adiabatic index $\gamma$ is used in lieu of the energy equation.
Manipulating the equations above and using $p\rho^{-\gamma} = constant$, one can
find the explicit relations for the wave amplitude and propagation speed:
\begin{equation}
    \begin{matrix}
        P_A = c_s^2\rho_A \\
        \mathbf{u_A} = \frac{\mathbf{k}}{\left|\mathbf{k}\right|} c_s \frac{\rho_A}{\rho_0}, \\
        \end{matrix}
\end{equation}\label{eq:acoustic}
where $c_s \equiv \sqrt{\gamma P_0 / \rho_0}$ is the wave propagation speed,
commonly known as the sound speed.

Finally, one can show from the characteristic equation that both the pressure
and velocity waves will have two modes traveling in opposite directions and the
density wave having a third, stationary mode.

\subsection{Performance and scalability}
\label{sec:sub:perf}

\begin{figure}[ht!]
    \centering
    \begin{tabular}{cc}
        \includegraphics[width=.48\textwidth]
        {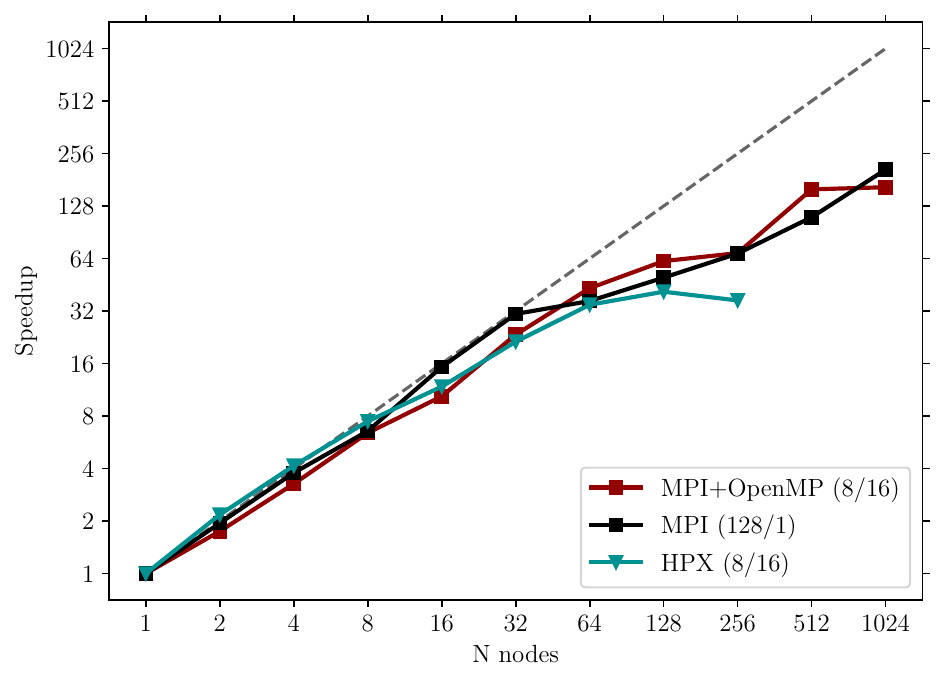} &
        \includegraphics[width=.48\textwidth]
        {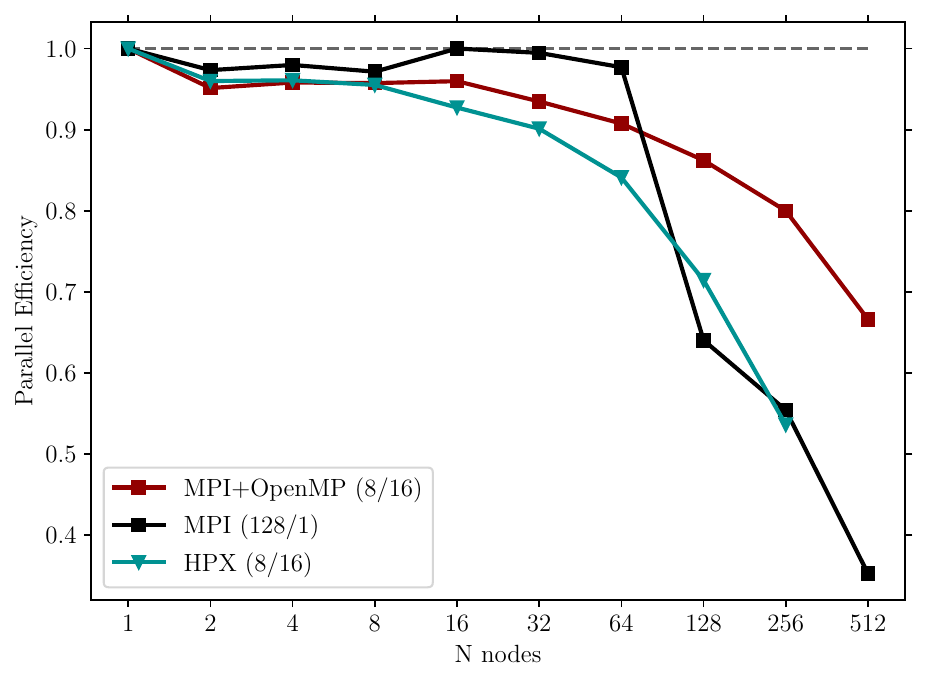}
    \end{tabular}
    \caption{Performance of the 3D radiation benchmark on Chicoma. 
    The notation (128/1) denotes the run configuration as (\#processes/\#MPI ranks).
    Left: strong scaling on 1024 nodes shows near-ideal speedup for MPI and MPI+OpenMP (dashed line = ideal linear scaling).
    Right: weak scaling efficiency, with MPI+OpenMP maintaining 70\% on 512 nodes.}
    \label{fig:scaling}
\end{figure}

We perform our scaling tests on LANL Chicoma supercomputer. Chicoma’s standard partition has
1792 nodes; each node contains two AMD EPYC Rome 64-core CPUs (7H12 @ 2.6 GHz)
with 512 GB of memory (16–32 GiB DIMMs). The GPU partition, although not used in
the scaling tests presented in Figure~\ref{fig:scaling}, consists of 118 nodes
equipped with AMD EPYC 7713 processors and NVIDIA A100 GPUs.

The scaling experiments in Figure~\ref{fig:scaling} demonstrate that HARD
achieves excellent parallel efficiency on Chicoma. For strong scaling, the 3D
radiation benchmark maintains near-ideal speedup up to thousands of cores,
confirming that the task-based decomposition in \flecsi combined with Kokkos
node-level parallelism enables effective utilization of large CPU partitions.
The weak scaling results further show that performance remains consistent as
both problem size and node count increase proportionally, indicating that HARD
sustains high throughput without degradation when tackling ever-larger
workloads.

In addition to these CPU-focused results, we have begun exploring GPU
acceleration on Chicoma’s A100 nodes. Preliminary tests show that a single GPU
achieves a 7× speedup compared to a CPU node, with performance reaching 67.6
million cell updates per second. These results highlight the strong potential of
HARD on GPU architectures, and ongoing work is focused on scaling across
multiple GPUs to fully characterize performance and portability in heterogeneous
environments.

Overall, these results validate the design goal of performance portability
across both CPU- and GPU-based systems. They demonstrate that HARD can not only
exploit existing large-scale clusters efficiently, but also serve as a scalable
foundation for even larger simulations expected on forthcoming exascale-class
machines.

\section{Impact}
\label{sec:impact}

\hard enables performance-portable simulations that are optimized for modern
heterogeneous architectures and extreme-scale parallelism. Our demonstration
problems highlight \hard's ability to accurately and robustly handle a wide
variety of physical scenarios, ranging from canonical test problems to
application-inspired benchmarks. These position \hard as not only a validation
platform for new algorithms, but also a research tool for investigating the
fundamental behavior of RHD systems. While additional physics modules may be
required to model the full complexity of HEDP phenomena, \hard already serves as
a versatile and reliable testbed for exploring a broad class of HEDP problems.
Its modularity and scalability position it as a valuable tool for advancing both
the development and validation of next-generation HEDP simulation capabilities.
Beyond HEDP, \hard's generality also makes it a valuable tool for astrophysical
simulations, laboratory plasma experiments, and other domains where
radiation–matter interactions play a central role.

\section{Conclusions}
\label{sec:conclusion}
We have presented \hard, a performance-portable RHD code
built on the \flecsi framework and designed to exploit the capabilities of
modern heterogeneous computing architectures. Through a series of benchmark test
problems, we have demonstrated the code’s flexibility, accuracy, and usability
across a range of relevant scenarios. Looking ahead, we plan to expand \hard by
incorporating advanced radiation transport models, including multi-group
approaches and higher-order diffusion schemes, thereby enhancing its ability to
capture spectral effects and anisotropic transport phenomena. The planned
integration of \flecsolve will further enable the efficient solution of
large-scale implicit radiation systems, a key capability for simulating
realistic, high-fidelity HEDP experiments. With these enhancements, \hard is
poised to become a robust and accurate tool for tackling complex RHD problems
across diverse disciplines, including HEDP and astrophysics.

\section*{Acknowledgements}
The \hard project is supported by the Los Alamos National Laboratory (LANL)
Advanced Simulation and Computing Program. LANL is operated by Triad National
Security, LLC, for the National Nuclear Security Administration of the U.S.
Department of Energy (contract no. 89233218CNA000001). This work is authorized
for unlimited release under LA-UR-25-27266.

The work reported in this paper would not have been possible without close
collaborations with the \flecsi and Applications teams.

\bibliographystyle{elsarticle-num} 
\bibliography{softwareX}

\end{document}